\documentclass[aps, prd, twocolumn, showpacs, superscriptaddress, groupedaddress]{revtex4-1} 
\usepackage{graphicx}	
\usepackage{multirow}
\usepackage{amssymb}
\usepackage{dcolumn}
\usepackage{color}
\usepackage{subfigure, rotating, bm, array}
\usepackage[pagebackref=false, colorlinks=true]{hyperref}
\hypersetup{
linkcolor=blue,     
citecolor=blue,     
urlcolor=blue}      
\begin{document}
\title{Comparing Shadows of Blackhole and Naked Singularity}
\author{Kanwar Preet Kaur}
\email{kanwarpreet27@gmail.com}
\affiliation{International Center for Cosmology, Charusat University, Anand, GUJ 388421, India}

\author{Pankaj S. Joshi}
\email{psjprovost@charusat.ac.in}
\affiliation{International Center for Cosmology, Charusat University, Anand, GUJ 388421, India}

\author{Dipanjan Dey}
\email{dipanjandey.icc@charusat.ac.in}
\affiliation{International Center for Cosmology, Charusat University, Anand, GUJ 388421, India}

\author{Ashok B. Joshi}
\email{gen.rel.joshi@gmail.com}
\affiliation{International Center for Cosmology, Charusat University, Anand, GUJ 388421,  India}

\author{Rucha P. Desai}
\email{ruchadesai.neno@charusat.ac.in}
\affiliation{PDPIAS, Charusat University, Anand, GUJ 388421, India}

\date{\today}

\begin{abstract}
It is now theoretically well established that not only a black hole can cast shadow, but other compact objects such as naked singularities, gravastar or boson stars can also cast shadows. An intriguing fact that has emerged is that the event horizon and the photon sphere are not necessary for a shadow to form. Now, when two different types of equally massive compact objects cast shadows of same size, then it would be very difficult to distinguish them from each other. However, the nature of the nulllike and timelike geodesics around the two compact objects would be different, since their spacetime geometries are different. Therefore, the intensity distribution of light emitted by the accreting matter around the compact objects would also be different. In this paper, we emphasize this phenomenon in detail. Here, we show that a naked singularity spacetime, namely, the first type of Joshi-Malafarina-Narayan (JMN1) spacetime can be distinguishable from the Schwarzschild blackhole spacetime by the intensity distribution of light, though they have same mass and shadow size. We also use the image processing techniques here to show this difference, where we use the theoretical intensity data. The differences that we get by using the image processing technique may be treated as a theoretical template of intensity differences, which may be useful to analyse the observational data of the image of a compact object.
\\
\\\begin{large}
{\Large •}
\end{large}
 $\boldsymbol{key words}$ : Blackhole spacetime, Naked Singularity spacetime
\end{abstract}

\maketitle
\section{Introduction}
Recent observations of the shadow \cite{Akiyama:2019fyp} of the central ultra-compact object of M87 galaxy has drawn enormous attention to the subject of general relativistic lensing and the shadow formation. The observed shadow of the galactic center M87 provides us important information of physical (e.g. spin, mass, etc.) and causal structure (e.g. existence of event horizon) of the central compact object. The new generation Event Horizon Telescope (ngEHT) group is continuously trying to observe the shadow of our Milkyway galaxy center. On the other hand, GRAVITY and SINFONI groups are tracking the trajectories of `S'-stars around our galactic center Sgr-A* \cite{M87, Eisenhauer:2005cv, center1}. Those `S'-stars are very close to the Sgr-A* and therefore general relativistic effects on the trajectories of those stars are expected to be observed. The nature of the trajectories of `S'-stars along with the shadow and other dynamical phenomenon (e.g. properties of accretion disk, Lense-thirring effect on Neutron stars, etc.) around Sgr-A* will help us to understand the physical and causal structure of the same. In this context, there are many papers where authors compare the shadow of a black hole with the shadow of other compact objects (e.g. naked singularities, gravastar, boson star, etc.) \cite{Shaikh:2019hbm,Gralla:2019xty,Abdikamalov:2019ztb,Yan:2019etp,Vagnozzi:2019apd,Gyulchev:2019tvk,Shaikh:2019fpu,Dey:2013yga,Dey+15,Shaikh:2018lcc,Joshi2020,Paul2020,Dey:2020haf,Dey:2020bgo,abdujabbarov_2015a, atamurotov_2015, abdujabbarov_2017, abdujabbarov_2015b, younsi_2016, papnoi_2014, bambi_2013a, ohgami_2015, stuchlik_2018, stuchlik_2019}. In order to analyse the astrometric data of `S' stars' trajectories, detailed theoretical studies on timelike orbits around different types of compact object have been done in \cite{Martinez:2019nor, Eva,Eva1,Eva2,tsirulev,Joshi:2019rdo,Bhattacharya:2017chr,Bambhaniya:2019pbr,Dey:2019fpv,Bam2020,Lin:2021noq,Deng:2020yfm,Deng:2020hxw,Gao:2020wjz,aa4}.

It is a general belief that a shadow is cast by black hole only \cite{Kormendy}. Therefore, the image of the shadow of M87 galaxy center is considered to be the shadow of a central super massive blackhole. However, in many literature, it is shown that shadow is not only the property of a black hole, but it can also be cast by various horizonless compact objects \cite{Shaikh:2019hbm,Gyulchev:2019tvk,Shaikh:2019fpu,Dey:2013yga,Dey+15,Shaikh:2018lcc,Joshi2020,Paul2020,Dey:2020haf,Dey:2020bgo}. One of the important horizonless compact objects is the naked singularity. Though the cosmic censorship conjecture (CCC) does not allow horizonless strong singularity \cite{penrose}, there are series of literature where it is shown that naked singularities can be formed during the continual gravitational collapse of an inhomogeneous matter cloud \cite{eardley, christodoulou, joshi, goswami, joshi2, vaz, jhingan0, deshingkar, mena, magli1, magli2, giambo1, harada1, harada2, joshi7, mosani1, mosani2, mosani3, mosani4, mosani5}. In \cite{Joshi:2011zm}, it is shown that a non-zero tangential pressure may resist the formation of trapped surfaces around the central high density region of the collapsing matter cloud which causes the formation of central naked singularity in large co-moving time. There are many literature where the properties of the shadow cast by naked singularities are extensively discussed \cite{Shaikh:2019hbm,Gyulchev:2019tvk,Shaikh:2019fpu,Dey:2013yga,Dey+15,Shaikh:2018lcc,Joshi2020,Paul2020,Dey:2020haf,Dey:2020bgo}. From these studies it can be understood that the existence of the upper bound of the effective potential of null-geodesics causes the formation of a shadow. If the effective potential of null-geodesics of a spacetime has an upper bound then that spacetime can cast a shadow. The presence of the event horizon and photon sphere around a singularity are not necessary for the existence of a shadow.  In \cite{Joshi2020,Dey:2020bgo}, it is shown that nulllike and timelike naked singularities can cast shadow in the absence of a photon sphere. 

In these papers, authors show that the shadow of a nulllike naked singularity is smaller than the equally massive Schwarzschild blackhole's shadow. On the other hand, timelike naked singularity can cast shadow of size greater or less than the size of shadow of a Schwarzschild black hole with equal Schwarzschild mass. In \cite{Dey:2020haf}, authors describe a scenario where a thin matter shell exists at the junction of two spacetimes and due to the thin shell, the spacetime structure can cast shadow in the absence of photon sphere. In all the above mentioned scenarios, the spin of the central compact objects are considered to be zero. Inclusion of spin can make a spacetime more realistic and the shadow cast by that spacetime can give us more realistic shadow properties of a compact object. In \cite{Bardeen:1973tla,Hioki:2009na,Johannsen Psaltis, Bambi:2011ek, Hansen:2013owa,Rezzolla:2014mua,Yagi:2016jml,Yunes:2011we,aa2}, shadows cast by various rotating black holes are discussed. It can be shown that   the upper bound of the effective potential of null geodesics does not exist in the Kerr naked singularity spacetime and therefore, it cannot cast shadow \cite{Bambhaniya:2021ybs}. However, in \cite{Bambhaniya:2021ybs}, it is shown that deformation in Kerr spacetime can make the scenario different. A deformed Kerr spactime can cast shadow, even though the central strong singularity is directionally naked. These novel features of various naked singularities might be very useful to understand the observed shadow of a compact object.

There are several theoretical models which predict that shadow size of a compact object can be same as the size of the equally massive Schwarzschild black hole's shadow \cite{Shaikh:2018lcc}. Therefore, it is very difficult to understand the causal structure of the compact object using the shadow size. When two equally massive compact objects having different causal structures cast shadows of same radius, then one needs to predict other types of possible physical signatures which can be used to distinguish one of the compact objects from other. The thermal property of the accretion disk around compact objects is one of the important properties of a compact object which can be used to understand the causal structure of the same. In \cite{Guo:2020tgv}, it is shown that the thermal properties of the accretion disk around a black hole can be significantly different from those for an equally massive naked singularity. Orbital dynamics of stars around a compact object also can be used to predict the causal nature of that compact object. In \cite{Bambhaniya:2019pbr, Dey:2019fpv, Dey:2020haf}, orbital dynamics of a test particle around different types of compact objects are discussed in detail. It is shown that the orbital dynamics around a naked singularity can be significantly distinguishable from that for an equally massive Schwarzschild black hole.

In this paper, we propose a new method to distinguish a black hole from a naked singularity spacetime having similar size of shadow and mass. In the asymptotic observer's sky, the image of a compact object may consist of a central dark region surrounded by a bright region where at the boundary of the dark region, the intensity of light is maximum and decreases gradually as radial distance from the central compact object increases. The central dark region, in the distant observer's sky, is known as the shadow of the compact object. From the accreting matter around a compact object, the light geodesics with different impact parameters reach the asymptotic observer. Due to the existence of an upper bound of the effective potential of null geodesics, in the shadow region, the intensity of light is significantly lower than that for the region just outside of the shadow. Since two different compact objects having equal mass and shadow size can have different geometry, the intensity of light inside the shadow region  would not be similar. Therefore, the behaviour of the intensity of light inside the shadow region can give us  important informations regarding the causal structure of the central compact object.

In the present work, we investigate the intensity distribution of light inside the shadow region for a naked singularity spacetime, namely Joshi-Malafarina-Narayan-1 (JMN1) spacetime and we compare it with the intensity distribution of light inside the shadow region of an equally massive Schwarzschild black hole.
In this paper, the shadow images of Schwarzschild black hole and JMN1 naked singularity are built from MATLAB using the theoretical intensity data which are acquired from their respective spacetimes. Herein, the image of the Schwarzschild blackhole's shadow is compared with the shadow cast by JMN1 naked singularity. Image processing tool is adopted to provide an additional in-depth comparative analysis of the massive object shadow images. In doing so, three significant image properties, i.e., intensities, contrast and structure are considered. Based on these
properties, shadow images are compared using some of the standard similarity
metrics of template matching techniques in image processing. These metrics would widen the scope for  determining the characteristics with which the black hole shadow image differs or matches the naked singularities shadow images. Furthermore, difference between the shadow images is depicted using histogram plot. An image histogram is a graphical representation of the pixel intensity distribution over number of pixels. In this analysis, histogram is plotted to observe any minute change in the pixel intensity. As we have stated above, we use the theoretical data for the image processing technique. However, using this technique, one can analyse the observational data of the shadow of a compact object. The distinguishable features of naked singularities' shadow which is described by the image processing technique can be observationally very much important to analyse the shadow and the causal structure of a compact object.


 \begin{figure*}
 \centering
 \subfigure[]
 {\includegraphics[width=76mm]{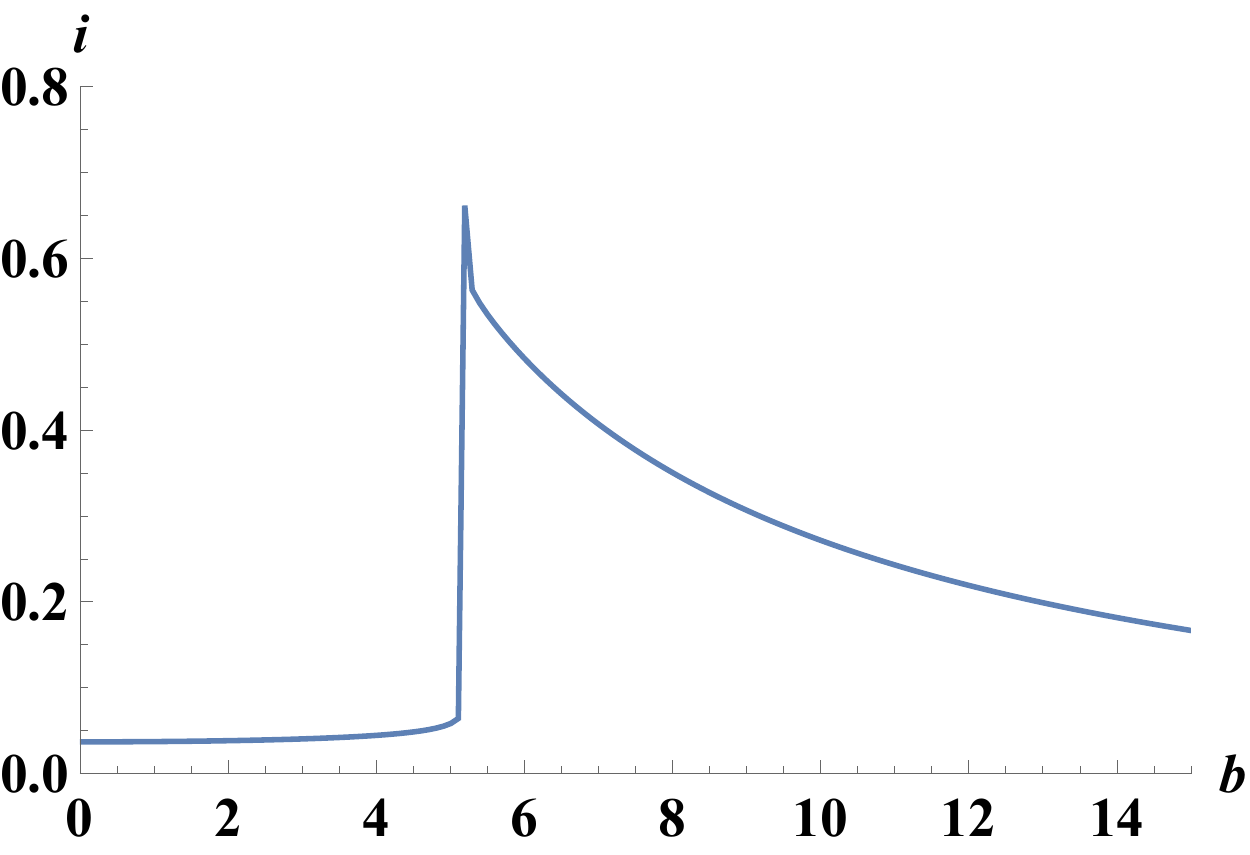}\label{fig1in1}}
 \hspace{0.2cm}
 \subfigure[]
 {\includegraphics[width=76mm]{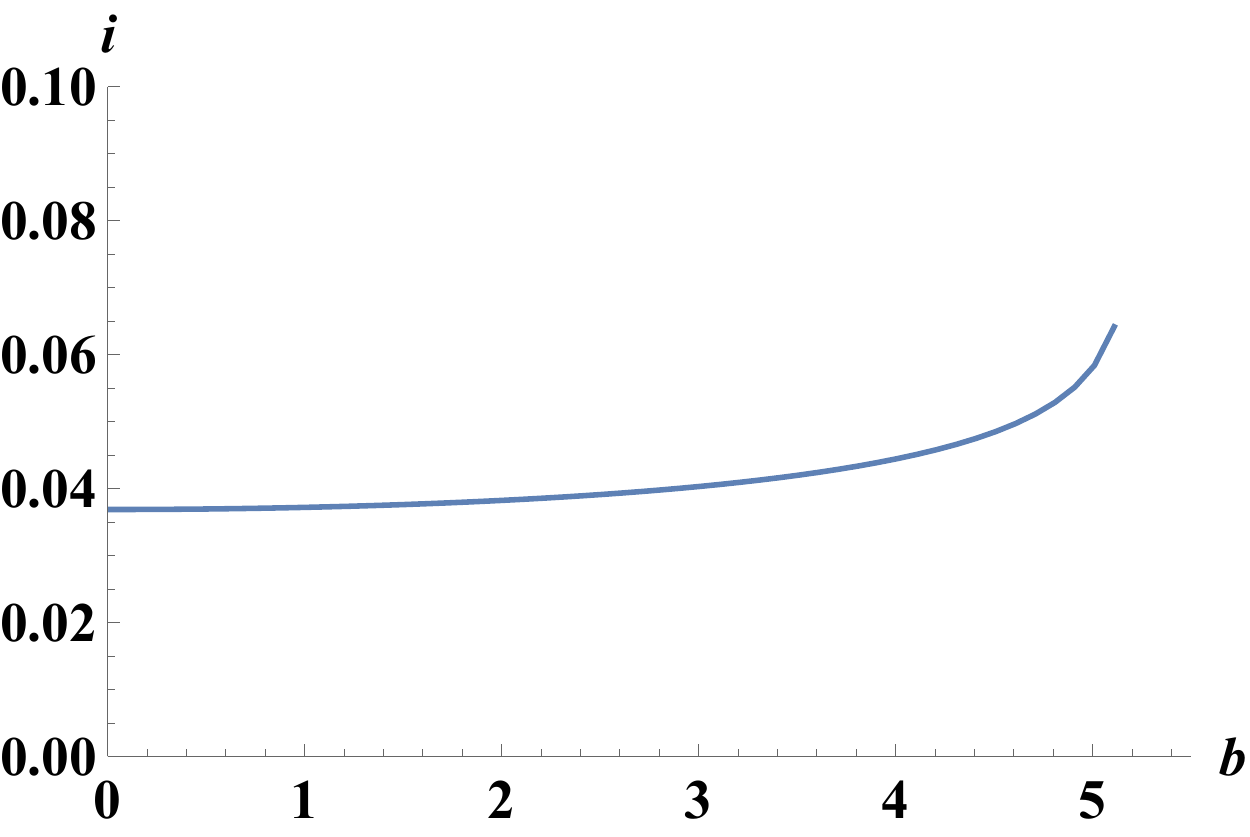}\label{fig1in2}}\\
 \hspace{0.2cm}
 \subfigure[]
 {\includegraphics[width=76mm]{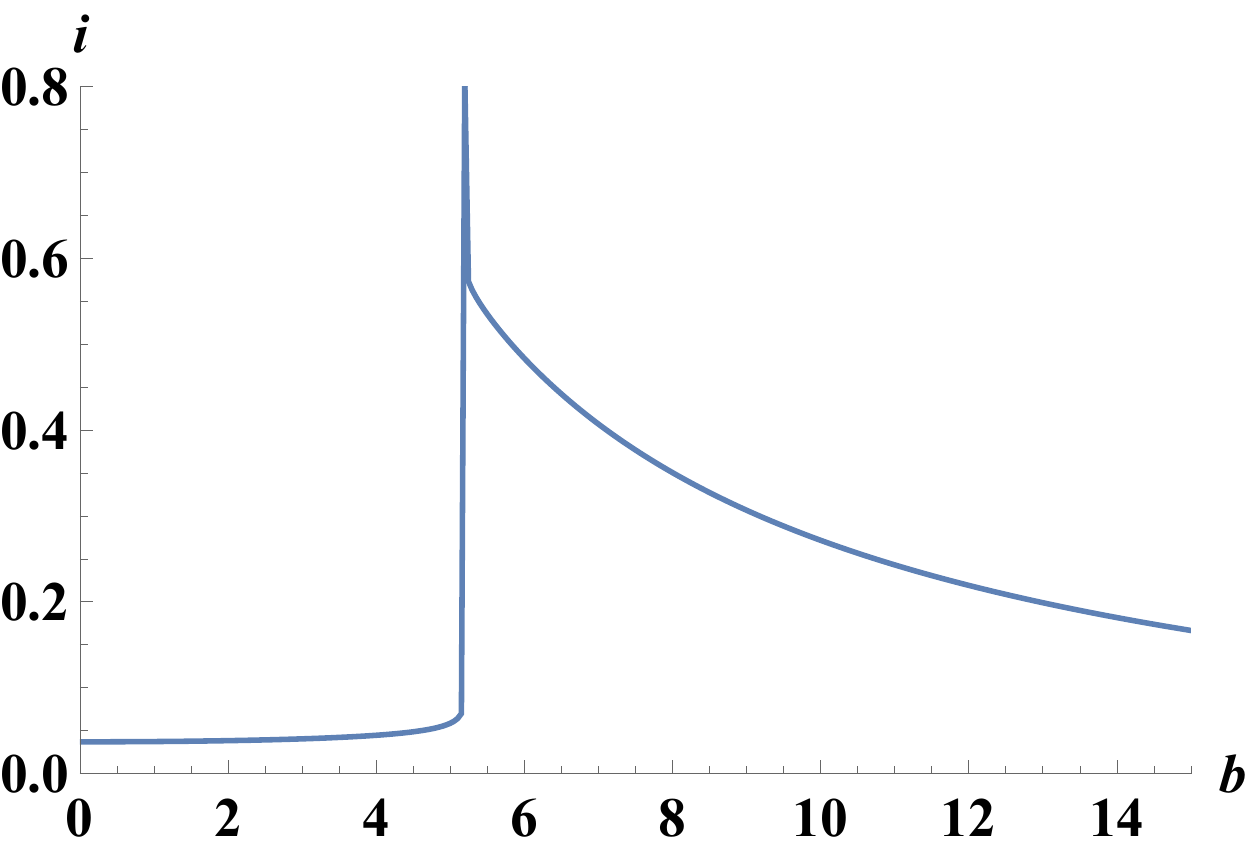}\label{fig1in3}}
 \hspace{0.2cm}
 \subfigure[]
 {\includegraphics[width=76mm]{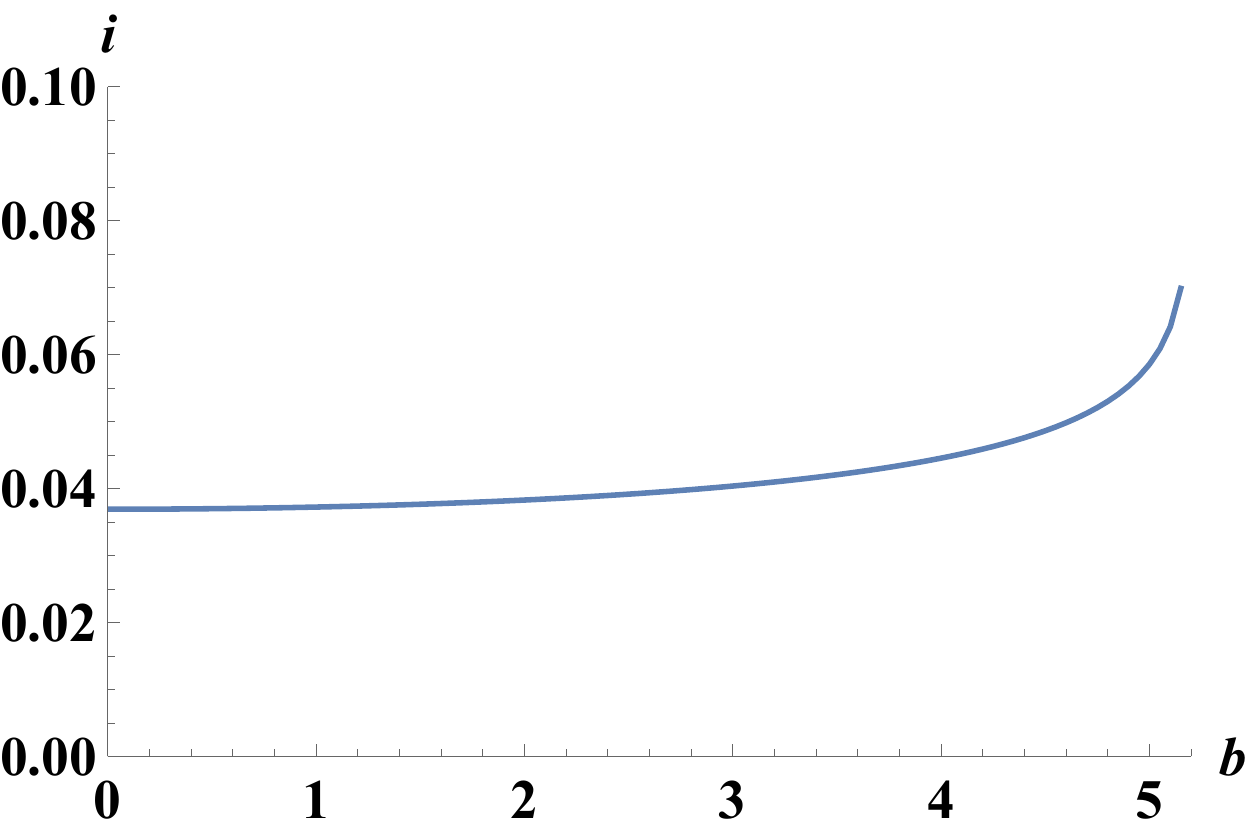} 
 \label{fig1in4}}\\
 \caption{Figs.~(\ref{fig1in1}), (\ref{fig1in3}) show  how the intensity distribution of light changes with impact parameter $b$ in JMN1 (with $M_0=0.7$) and Schwarzschild spacetimes respectively. On the other hand, the Figs.~(\ref{fig1in2}), (\ref{fig1in4}) show the intensity ditribution up to the shadow boundary of JMN1 and Schwarzschild spacetimes respectively.}\label{fig1}
\end{figure*}

The plan of the paper is as follows. In section (\ref{sectionthree}), we discuss the nature of the light trajectories in JMN1 spacetime. In the section (\ref{sectionfour}), we show the comparative study of the intensity distribution between JMN1 and  Schwartzschild spacetimes. In section (\ref{imageprocess}), we depict qualitative image difference using image processing and the numerical data analysis. Finally, in section (\ref{result}), we discuss the results and conclusion of the work. Throughout the paper we consider $G = c = 1$.

\section{Shadow and Light Trajectories in JMN1 naked singularity spacetime}\label{sectionthree}

\subsection{Null geodesics and shadow in a spherically symmetric static spacetime}
In this section, we discuss the general solution of null geodesics around a compact object, and investigate the contributions of different regions of a spacetime to the light's intensity distribution inside and outside the shadow. To start with, we consider the following spherically symmetric, static spacetime,
\begin{equation}
    ds^2 = - g_{tt}dt^2 + g_{rr}dr^2 + S(r)  (g_{\theta \theta} d\theta^2 + g_{\phi \phi} d\phi^2)\,\,,
    \label{static}
\end{equation}
where $g_{tt}$,~$g_{rr}$ are the functions of r only, and the azimuthal part of the spacetime shows the spherical symmetry.
As we know, the conserved angular momentum ($\mathbb{L}$) and energy ($\mathbb{E}$) per unit rest mass of a freely falling particle in the the above mentioned spherically symmetric and static spacetime are,
\begin{eqnarray}
\mathbb{L} = S(r) g_{\phi \phi}\frac{d\phi}{d\lambda}\, ,\,\,\nonumber\\
\mathbb{E} =  g_{tt}\frac{dt}{d\lambda}\,\,,
\label{conservedJNW}
\end{eqnarray}
where $\lambda $ is the affine parameter. Through out this paper, we only consider geodesic motion on equatorial plane. Therefore, using null condition $K^{\mu} K_{\mu} = 0$ where $K^{\mu}$ is the four-velocity of photon, we get 
\begin{equation}
    \frac{1}{b^2}=\frac{g_{tt} g_{rr}}{h^2} \left(\frac{dr}{d\lambda}\right)^2 + W_{eff}\,\, ,
\end{equation}
where $W_{eff} = g_{tt}/(S(r) r^2)$ and $1/b^2 = \mathbb{E}^2 / \mathbb{L}^2$. Parameter $b$ is known as the impact parameter of light geodesic which can be calculated from the conserved energy and angular momentum and $W_{eff}$ is the effective potential of null geodesics. As we know, in the strong gravity region in a spherically symmetric spacetime, there may exist a spherical surface which consists of unstable circular null geodesics and that surface is known as photon sphere. Therefore, photon sphere exists when at a particular radius ($r_{ph}$), the following properties of effective potential hold,
 $$\frac{dW_{eff}}{dr}|_{r_{ph}} = 0\,\, , \frac{d^2W_{eff}}{dr^2}|_{r_{ph}} <0\,\, .$$
There are some scenarios where stable circular orbits of null geodesic exist \cite{Bhattacharya:2017chr, Paul2020}. The spherical surface of the stable circular null orbits is known as anti-photon sphere. Null geodesics from a distance source with the impact parameter $b_{ph}$ corresponding to the photon sphere radius $r_{ph}$ may undergo multiple circular windings of radius $r_{ph}$ before plunging into the center or escaping away to infinity. On the other hand, photons coming from a distance source with the impact parameter $b > b_{ph}$ can not reach the photon sphere and escape away to the asymptotic observer. At the turning points of the scattered null geodesics, $\frac{dr}{d\lambda}|_{r_{tp}}=0$ and the corresponding impact parameter ($b_{tp}$) can be written as,
\begin{equation}
    b_{tp} = r_{tp} \sqrt{\frac{S(r_{tp})}{g_{tt}(r_{tp})}}\,\, ,
\end{equation}
where $r_{tp}$ is the radial distance of the turning point.  However, photons with impact parameter $b<b_{ph}$ are plunged into the central object and therefore, can not reach the distant observer. Hence, this phenomenon creates a shadow of radius $b_{ph}$ in the observer's sky. 

\subsection{Shadow cast by JMN1 spacetime}
\subsubsection{JMN1 spacetime} The JMN1 spactime is a spherically symmetric and static solution of Einstein field equation and it has a central naked singularity \cite{Joshi:2011zm}. In \cite{Joshi:2011zm}, authors show that JMN1 spacetime can be thought to be the equilibrium end state of the gravitational collapse of a matter cloud having zero radial pressure and non-zero azimuthal pressure. Due to the existence of the non-zero tangential pressure inside the collapsing matter, the collapsing matter cloud equilibrates in a large co-moving time and that non-zero pressure resists the formation of trapped surfaces inside the matter cloud. Therefore, the central singularity formed in a large co-moving time is naked and the asymptotic spacetime can be shown as JMN1 spacetime. The line element of the JMN1 spacetime can be written as,
\begin{eqnarray}
 ds^2=-(1-M_0) \left(\frac{r}{R_b}\right)^\frac{M_0}{1-M_0}dt^2 + \frac{dr^2}{1-M_0} + r^2d\Omega^2\,\, ,
\label{JMN-1metric}
\end{eqnarray}
where $ 0 <  M_{0} < 1$, $d\Omega^2=d\theta^2+\sin^2\theta d\phi^2$ and $R_b$ is the boundary of JMN1 spacetime where it smoothly matches with external Schwarzschild spacetime. The external Schwarzschild spacetime can be written as,
\begin{eqnarray}
 ds^2=-\left(1-\frac{M_0 Rb}{r}\right) dt^2 + \frac{dr^2}{\left(1-\frac{M_0 Rb}{r}\right)} + r^2d\Omega^2\,\, .
\label{SCHmetric}
\end{eqnarray}
Therefore, the Schwarzschild mass of the internal JMN1 spacetime is $M=\frac{M_0 R_b}{2}$. 

From the Einstein field equations, we can get the following expressions of energy density and pressures,
\begin{equation}
 \rho = \frac{M_{0}}{r^2}\,\, , 
 P_{\theta} =  \frac{M_{0} \rho}{4(1 - M_{0})}\,\, ,P_r = 0\,\, .\label{eq17}
\end{equation}
Therefore, equation of state for anisotropic fluid of JMN1 spacetime can be written as $M_{0}/(6(1-M_{0}))$.

\begin{figure*}
\centering
{\includegraphics[width=90mm]{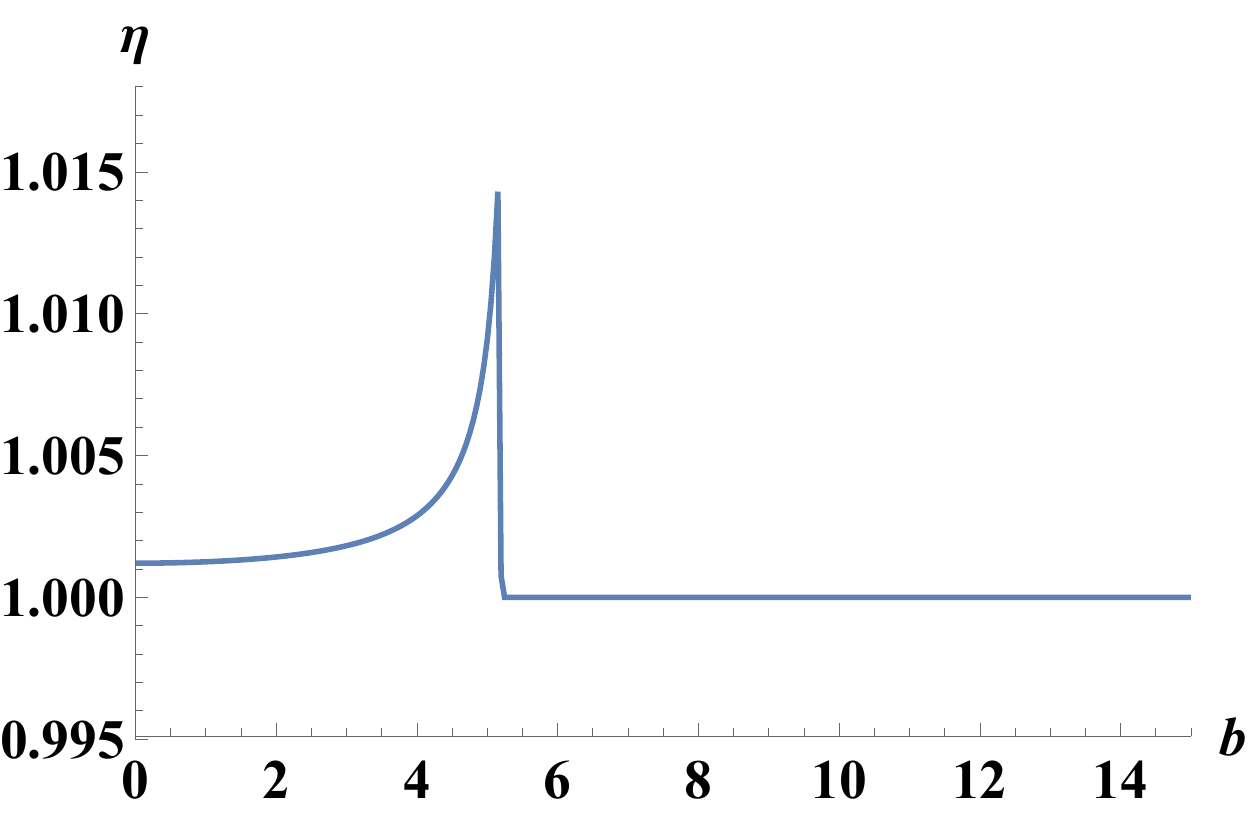}}
 \caption{In this figure, we show  how the ratio of the intensity ($\eta$) varies with impact parameter ($b$). Detailed analysis of the figure is given in the text.}\label{fig3Eta}
\end{figure*}

\subsubsection{Shadow of JMN1 spacetime}

As we know, in Schwarzschild spacetime, photon sphere exists at $r=3M$. Therefore, it can be shown that the radius of the shadow cast by the Schwarzschild blackhole is $b_{ph}=3\sqrt3 M$. JMN1 spacetime does not have the upper-bound of the effective potential of the null geodesics. However, for the spacetime configuration which is internally JMN1 and externally Schwarzschild spacetime, the effective potential can have an upper-bound for some specific range of parameters' values.
As we discussed previously, JMN1 spacetime can be smoothly matched with an external Schwarzschild spacetime and the Schwarzschild mass of that spacetime configuration can be written as $M=\frac{M_0 R_b}{2}$. Therefore, it can be seen that for  $M_0>\frac23$, the matching radius $R_b < 3M$. Therefore, for $M_0> \frac23$, a photon sphere of radius $3M$ exists in the external Schwarzschild spacetime and the spacetime configuration can cast a shadow of radius $3\sqrt3 M$ which is similar to the radius of the shadow cast by Schwarzschild black hole.   On the other hand, since $M_0<\frac23$ implies $R_b > 3M$, no photon sphere exists in the spacetime configuration and a `full-moon' image of the compact object would be seen in the asymptotic observer's sky \cite{Shaikh:2018lcc}. Hence, for $M_0> \frac23$, JMN1 spacetime can cast similar shadow what a Schwarzschild black hole can cast. However, since inside the matching radius $R_b$ the vacuum does not exist, nature of null geodesics is also different and that difference can be observed in the intensity distribution inside the shadow region. In the next section, we comparatively study the intensity distribution of light inside the shadow region of JMN1 and Schwarzschild spacetime.

\section{Comparative study of intensity distribution in naked singularity and black hole spacetime }\label{sectionfour}
In order to discuss the intensity profile of light coming from the radiating accreting matter around a central compact object, one need to specify the four velocities of emitter and  observer. In this paper, we consider a stationary asymptotic observer and the emitter is radially freely falling. Therefore, the asymptotic observer's four velocity $u^\mu_o = (1,0,0,0)$ and the four velocity of radially freely falling emitter in the spherically symmetric static spacetime mentioned in Eq.~(\ref{static}), can be written as, $u^{\alpha}_e \equiv \left[\frac{1}{g_{tt}(r)},-\frac{\left(1-g_{tt}(r)\right)}{g_{tt}(r)g_{rr}(r)},0,0\right]$.

In \cite{Shaikh:2018lcc, bambi_2013a}, authors extensively discuss the intensity profile of the light geodesics coming from radially freely falling, spherically symmetric accreting matter around a compact object. The intensity of light seen by an asymptotic observer on the (X,Y) plane can be written as, 
\begin{equation}
    I_{\nu_o}(X,Y)= \int_{\gamma}^{} g^3 j(\nu_e) dl_{prop}\,\, ,\label{eq31}
\end{equation}
where $\nu_e$ is the emitted photon frequency  measured in the rest frame of the emitter and $\nu_o$ is the photon frequency measured in asymptotic observer's frame. In the above expression of intensity,  $g=\nu_o/\nu_e$ is the red-shift factor, $j(\nu_e)$ is the emissivity per unit volume (in the emitter frame) and $dl_{prop}$ is the infinitesimal proper length in the rest frame of emitter which can be written as $dl_{prop}=-k_\alpha u^\alpha_e d\lambda$, where $k^\mu$ is the null four velocity. 
The red-shift factor ($g$) can be written as, 
\begin{equation}
    g=\frac{k_\alpha u^\alpha_e}{k_\beta u^\beta_o}\,\, .
\end{equation}
Here, we take a simple case where matter is radially freely falling towards the central compact object and it emits monochromatic light, where the emissivity is
\begin{equation}
    j(\nu_e) \propto \frac{\delta(\nu_e-\nu_*)}{r^2}\,\, ,
\end{equation}
where $\delta$ is the Dirac delta function. The infinitesimal proper length ($dl_e$) can be written in terms of red-shift factor and four velocity of light as, 
\begin{equation}
    dl_{prop}=-\frac{k_t}{g k^r}dr.\label{eq37}
\end{equation}
Using Eq. (\ref{eq37}) and Eq. (\ref{eq31}), and integrating over light trajectories for all the observed frequencies, we get the observed photon flux
\begin{equation}
    I_o(X,Y) \propto - \int_{\gamma}^{} \frac{g^3 k_t dr}{r^2 k^r}\,\, ,
\end{equation}
where $I_o(X,Y)$ is intensity distribution in the (X, Y) plane.

In Figs.~(\ref{fig1in1}), (\ref{fig1in3}), we show how the intensity distribution changes with impact parameter $b$ for JMN1 (with $M_0=0.7$) and Schwarzschild spacetime respectively. In Figs.~(\ref{fig1in2}), (\ref{fig1in4}), we show the intensity distribution up to the shadow boundary in these spacetimes. It is apparent from those diagrams that the central shadow region has a non-zero intensity distribution which can give us the information about the central compact object. Here, we consider the Schwarzschild mass $M=1$ for both JMN1 and Schwarzschild spacetimes.

For comparative study of intensity distribution between Schwarzschild and JMN1 spacetimes, we introduce new parameter $\eta$ which can be written as,
\begin{equation}
    \eta = \frac{I_{0(JMN1)}}{I_{0(SCH)}}\,\, ,
\end{equation}
where $I_{0(JMN1)}$ and $I_{0(SCH)}$ are the observed intensities of light in JMN1 and Schwarzschild spacetimes at a particular value of impact parameter $b$. Therefore, $\eta$ is a function of $b$.
In Fig.~(\ref{fig3Eta}), we show how the parameter $\eta$ varies with the impact parameter ($b$). It can be seen from the Fig.~(\ref{fig3Eta}) that the value of $\eta$ is greater than one in the region $0\leq b< 3\sqrt3$ and it takes the highest value as $b$ tends to $3\sqrt3$. On the other hand, in the region $b>3\sqrt3$, the value of $\eta$ is one and it is constant in that region. Therefore, inside the shadow region (i.e., $0\leq b <3\sqrt3$), the intensity of light in JMN1 spacetime is greater than that for Schwarzschild spacetime, whereas, outside that region the intensity of light in both spacetimes are similar. The reason behind the difference in the intensity distributions between JMN1 and Schwarzschild spacetimes inside the shadow region is that in $0\leq r\leq r_b$, the geometry of the spacetime configuration mentioned in Eqs.~(\ref{JMN-1metric}, \ref{SCHmetric}) is not similar to the Schwarzschild spacetime. Since the spacetime configuration has a central naked singularity, light rays from the region $0\leq r\leq r_b$ can escape and reach to the asymptotic observer. However, in Schwarzschild spacetime, no light can escape from the region $0\leq r \leq 2$. Since the external spacetime of the spacetime configuration is Schwarzschild spacetime, $\eta =1$ for $b> 3\sqrt 3$.

From the above analysis, we can say that two different compact objects having same Schwarzschild mass and shadow size, can be distinguishable by intensity distribution of light inside the shadow region.
We know in near future EHT (Event Horizon Telescope) will release the shadow of the Sagittarius A* (Sgr-A*) and the intensity distribution inside the shadow region can reveal the causal structure of the Sgr-A*. In the next section, using image processing technique, we show the same difference between  JMN1 and Schwarzschild spacetimes in the context of light intensities seen by an asymptotic observer.

\begin{figure*}
\centering
\subfigure[]
{\includegraphics[width=65mm]{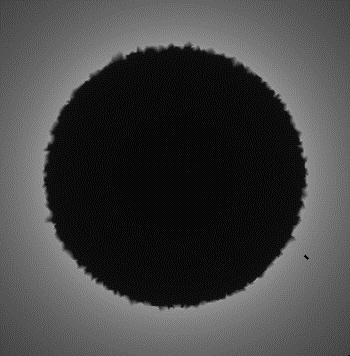}
\label{fig3in1}}
\hspace{0.05cm}
\subfigure[]
{\includegraphics[width=65mm]{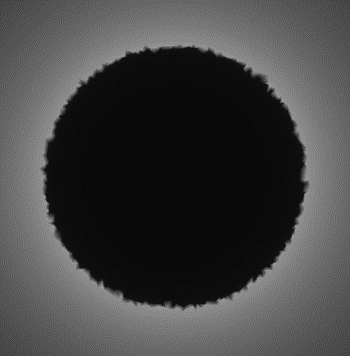}
\label{fig3in2}}\\
\caption[]{Shadow images created from 2000 interpolated intensity data points: Fig.~(\ref{fig3in1})~Schwarzschild Black Hole and Fig.~(\ref{fig3in2})~JMN1 Naked Singularity.}
\label{fig3}
\end{figure*}

\begin{figure*}
\centering
\subfigure[]
{\includegraphics[width=65mm]{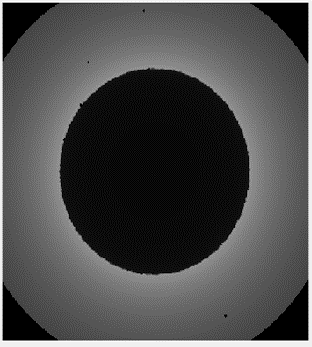}
\label{fig4in1}}
\hspace{0.05cm}
\subfigure[]
{\includegraphics[width=65mm]{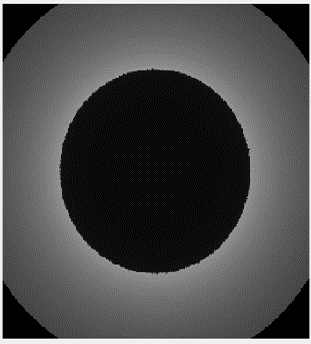}
\label{fig4in2}}\\
\caption[]{Shadow images created from 20,000 interpolated intensity data points: Fig.~(\ref{fig4in1})~Schwarzschild Black Hole and Fig.~(\ref{fig4in2})~JMN1 Naked Singularity.}
\label{fig4}
\end{figure*}

\begin{figure*}
\centering
\subfigure[]
{\includegraphics[width=55mm]{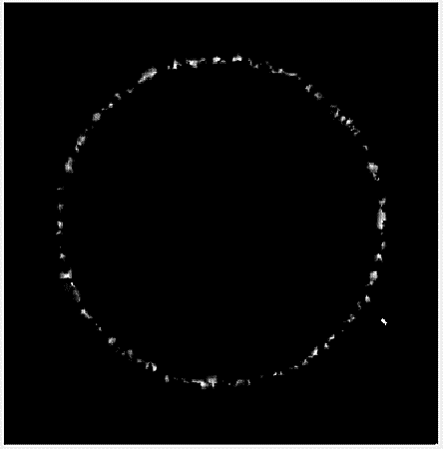}
\label{fig5in1}}
\hspace{0.05cm}
\subfigure[]
{\includegraphics[width=55mm]{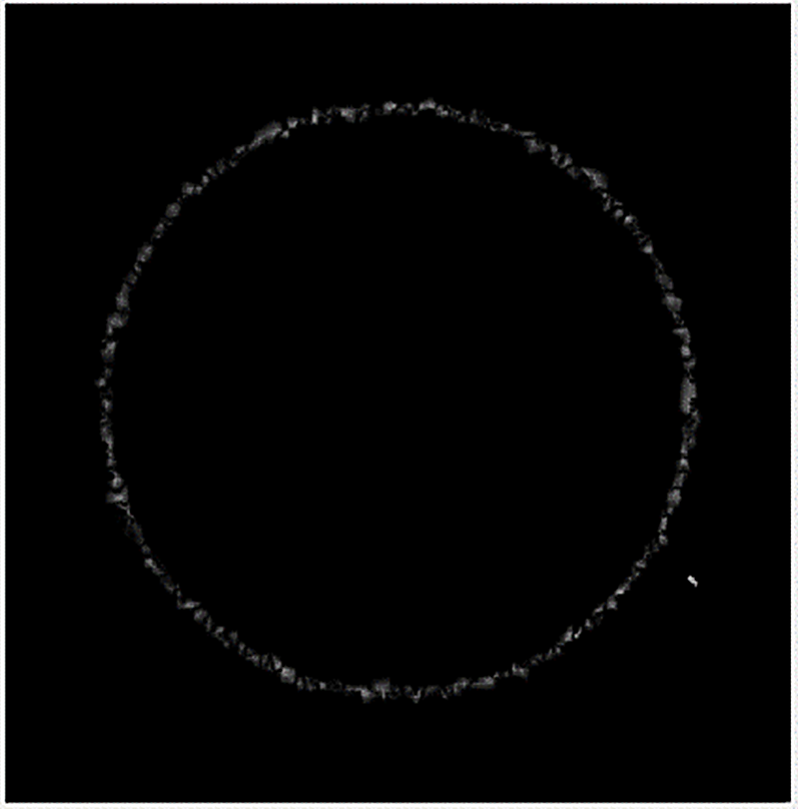}
\label{fig5in2}}
\hspace{0.05cm}
\subfigure[]
{\includegraphics[width=55mm]{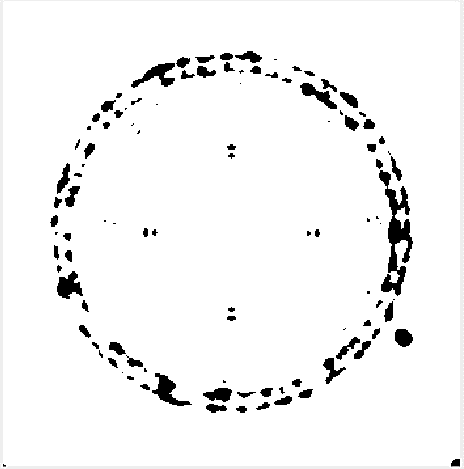}
\label{fig5in3}}\\
\caption[]{Difference and similarity between the Schwarzschild black hole and JMN1 naked singularity shadow images for CASE-I: Fig.~(\ref{fig5in1})~Arithmetic Difference Image, Fig.~(\ref{fig5in2}) ~Absolute Difference Image, and Fig.~(\ref{fig5in3})~Image based on SSIM index metric.}
\label{fig5}
\end{figure*}

\begin{figure*}
\centering
\subfigure[]
{\includegraphics[width=55mm]{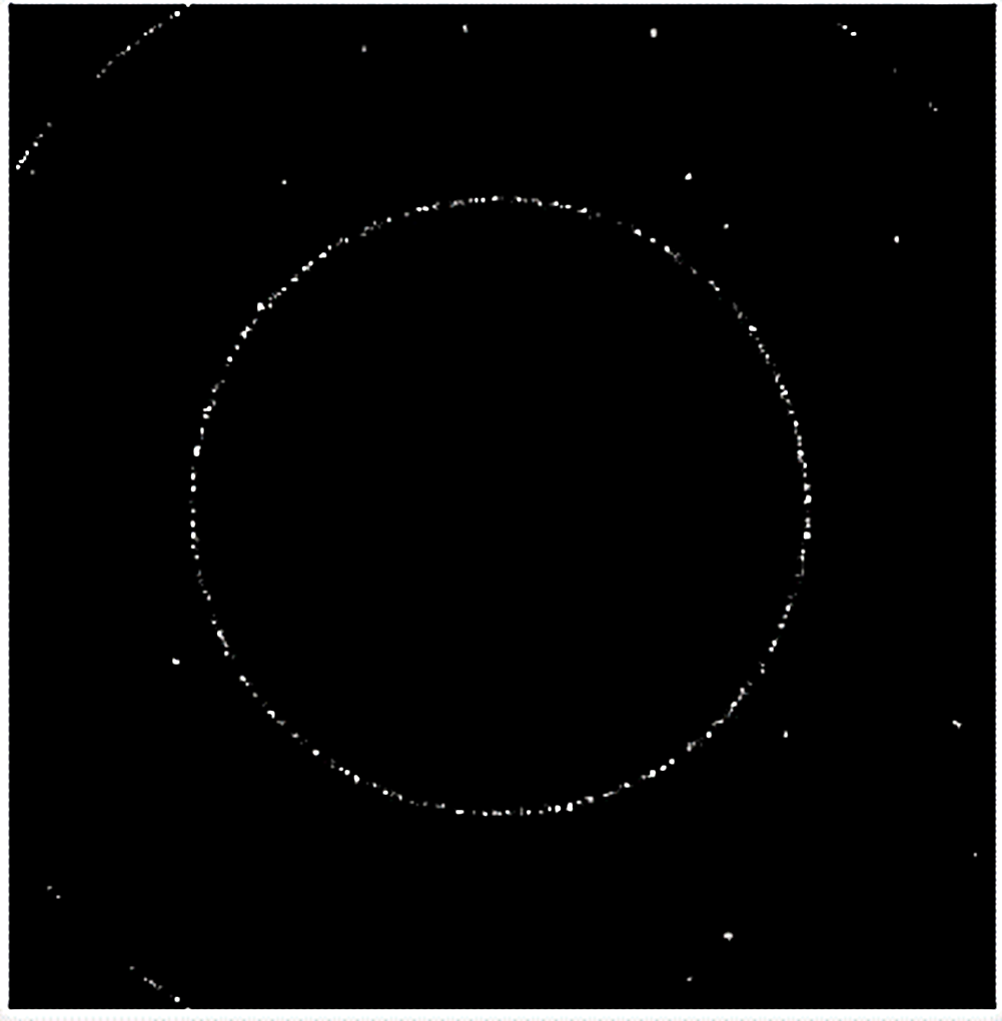}
\label{fig6in1}}
\hspace{0.05cm}
\subfigure[]
{\includegraphics[width=55mm]{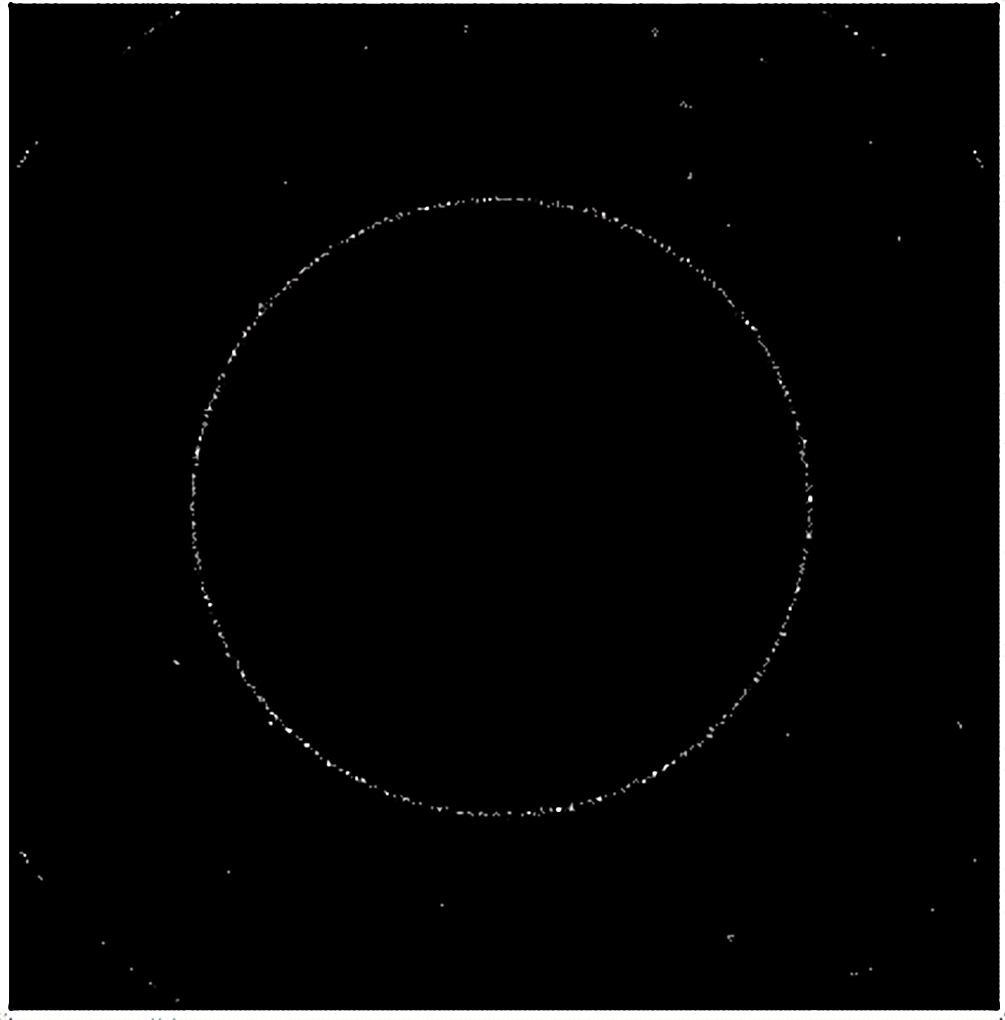}
\label{fig6in2}}
\hspace{0.05cm}
\subfigure[]
{\includegraphics[width=55mm]{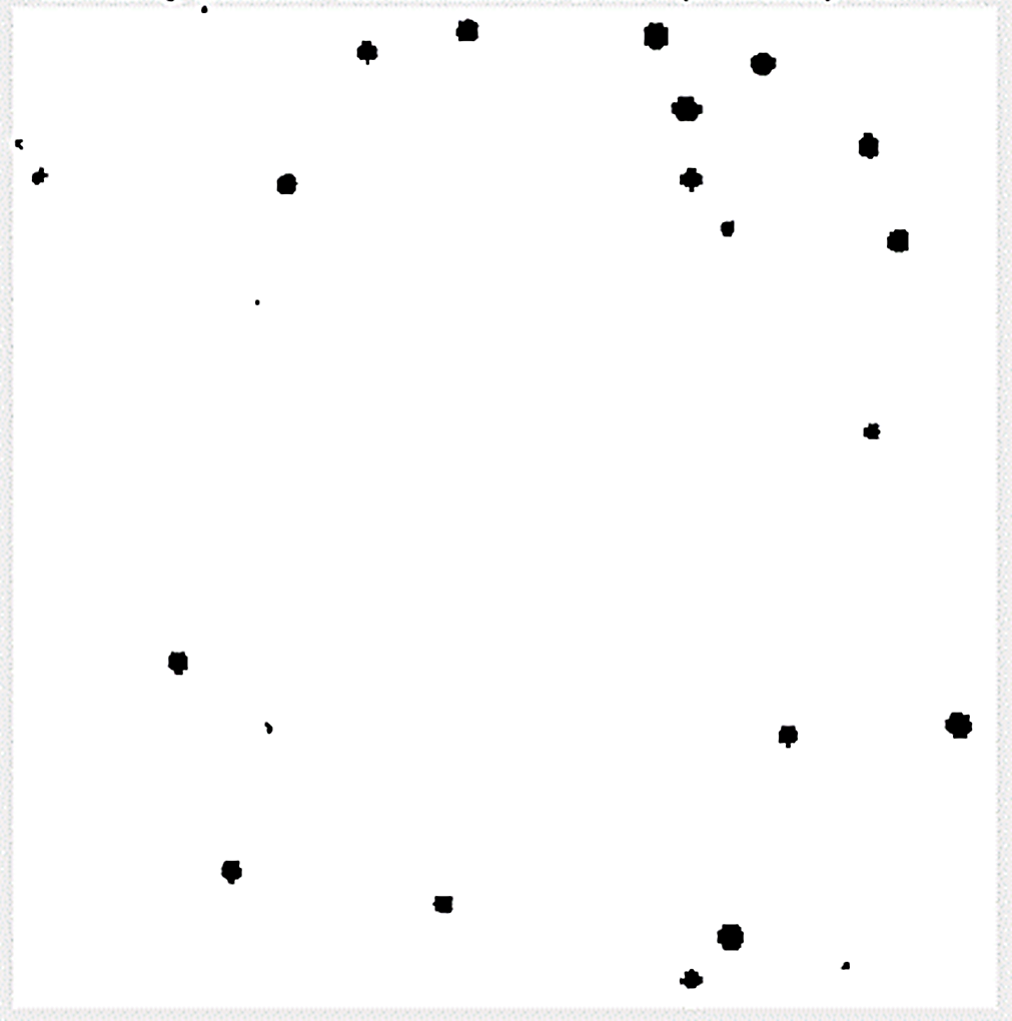}
\label{fig6in3}}\\
\caption[]{Difference and similarity between the Schwarzschild black hole and JMN1 naked singularity shadow images for CASE-II: Fig.~(\ref{fig6in1})~Arithmetic Difference Image, Fig.~(\ref{fig6in2}) ~Absolute Difference Image, and Fig.~(\ref{fig6in3})~Image based on SSIM index metric.}
\label{fig6}
\end{figure*}

\begin{figure*}
\centering
\subfigure[]
{\includegraphics[width=85mm]{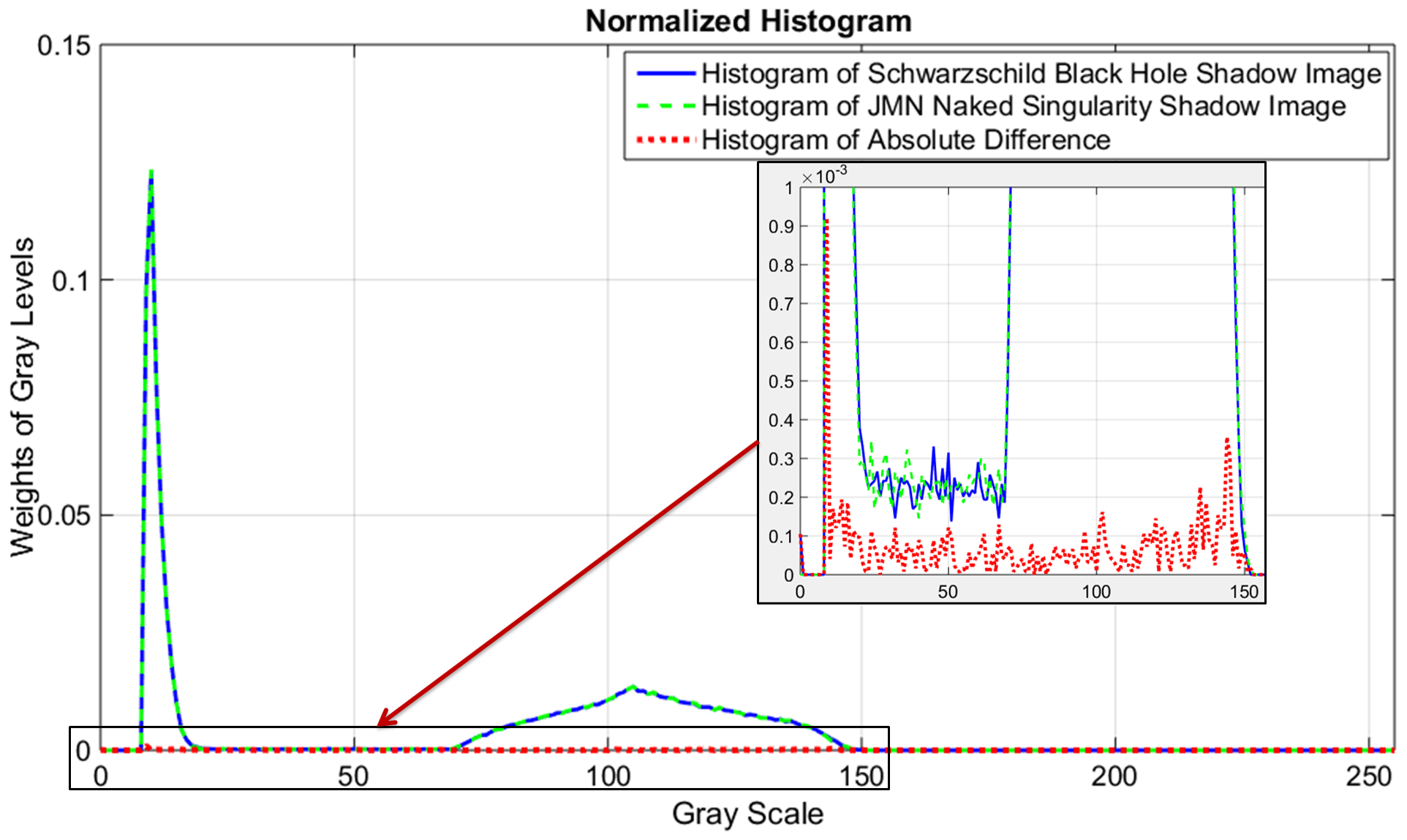}
\label{fig7in1}}
\hspace{0.05cm}
\subfigure[]{\includegraphics[width=85mm]{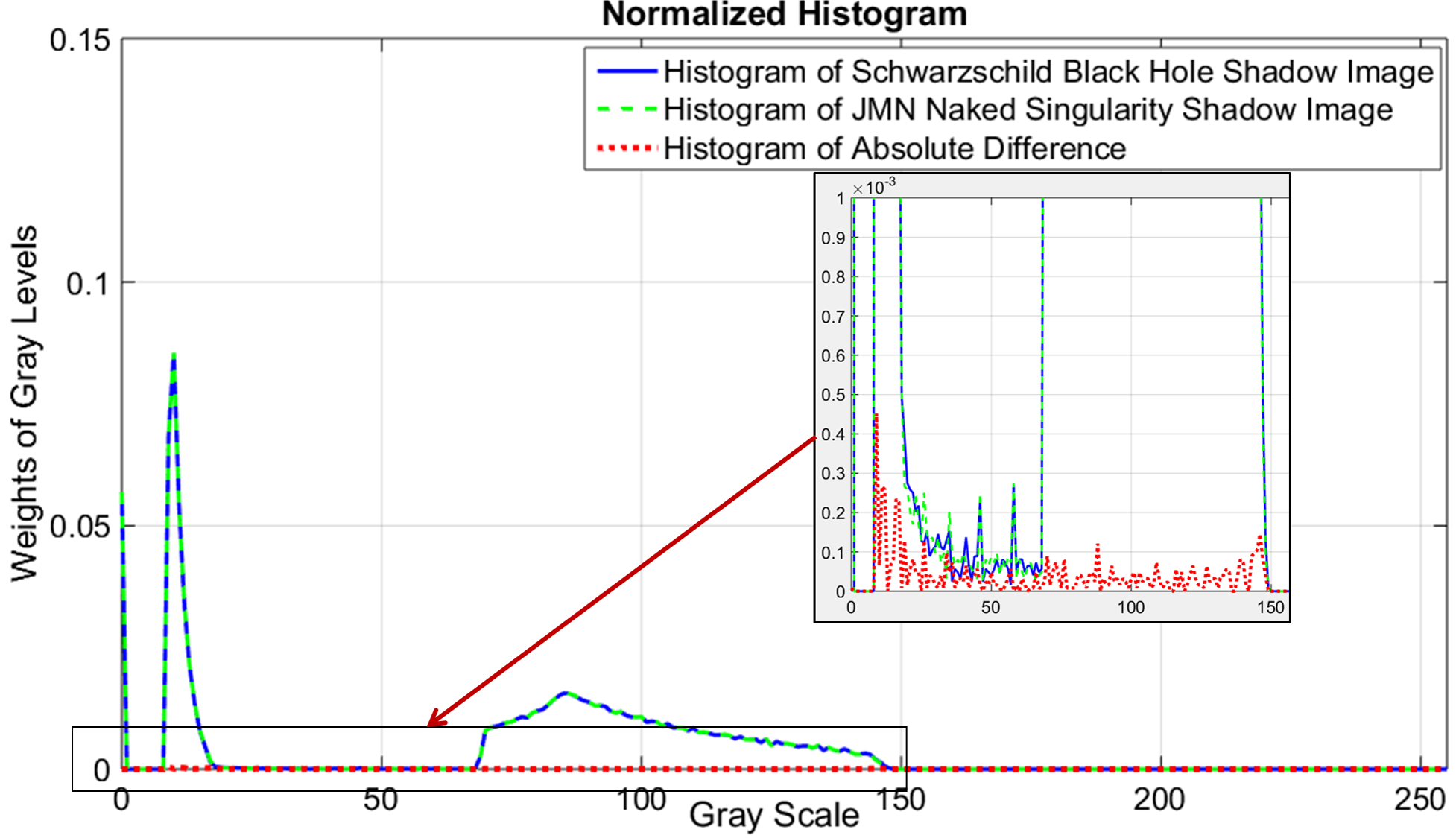}
\label{fig7in2}}\\
\caption[]{Normalized Histogram of absolute difference between the Schwarzschild black hole and JMN1 naked singularity shadow images  for Fig.~(\ref{fig7in1})~CASE-I and Fig.~(\ref{fig7in2})~CASE-II.}
\label{fig7}
\end{figure*}

\begin{table*}
  \centering
  \footnotesize
  \caption{Overall comparison between the Schwarzschild (SCH) black hole and JMN1 naked singularity shadow images}
    \begin{tabular}{p{6em}cccccccccc}
    \hline
    \hline
    \multirow{3}[0]{*}{\textbf{Parameters}} &
         \multicolumn{10}{p{43.5em}}{\centering{\textbf{Intensity Data Points to Interpolated points}}} \\
        \cline{2-11}
    \multicolumn{1}{c}{} & \multicolumn{1}{p{3.5em}}{\multirow{2}[0]{=}{\centering{\textbf{20,000 to 2000}}}} & \multicolumn{1}{p{3.5em}}{\multirow{2}[0]{=}{\centering{\textbf{20,000 to 20,000}}}} & \multicolumn{1}{p{3.5em}}{\multirow{2}[0]{=}{\centering{\textbf{20,000 to 20,000}}}} & \multicolumn{7}{p{28em}}{\centering{\textbf{2,00,000 to 20,000}}} \\
    \cline{5-11}
    \multicolumn{1}{c}{} &       &       &       & \multicolumn{1}{p{4em}}{\centering{\textbf{SCH(1) \& JMN1(1)}}} & \multicolumn{1}{p{4em}}{\centering{\textbf{SCH(2) \& JMN1(2)}}} & \multicolumn{1}{p{4em}}{\centering{\textbf{SCH(3) \& JMN1(3)}}} & \multicolumn{1}{p{4em}}{\centering{\textbf{SCH(4) \& JMN1(4)}}} & \multicolumn{1}{p{4em}}{\centering{\textbf{SCH(5) \& JMN1(5)}}} & \multicolumn{1}{p{4em}}{\centering{\textbf{SCH(6) \& JMN1(6)}}} & \multicolumn{1}{p{4em}}{\centering{\textbf{SCH\_avg \& JMN1\_avg}}} \\
    \hline
    \textbf{Percentage Arithmetic Difference Relative to Total Pixels} & 6.6532 & 6.7081 & 3.8860 & 1.5699 & 1.5420 & 1.5093 & 1.5408 & 1.5997 & 1.5610 & 1.8322 \\
    \hline
    \textbf{Correlation Coefficient } & 0.9962 & 0.9960 & 0.9820 & 0.9986 & 0.9988 & 0.9987 & 0.9987 & 0.9987 & 0.9988 & 0.9998 \\
    \hline
    \textbf{Percentage Absolute Difference Relative to Schwarzschild Image} & 1.7672 & 1.7830 & 4.3230 & 0.4767 & 0.4546 & 0.4533 & 0.4645 & 0.4779 & 0.4485 & 0.2261 \\
    \hline
    \textbf{Percentage Sum Square Difference } & 0.7691 & 0.7750 & 3.6770 & 0.2718 & 0.2450 & 0.2517 & 0.2614 & 0.2549 & 0.2366 & 0.0418 \\
    \hline
    \textbf{SSIM Index Value} & 0.9216 & 0.9113 & 0.9112 & 0.9713 & 0.9698 & 0.9722 & 0.9717 & 0.9698 & 0.9747 & 0.9777 \\
    \hline
    \textbf{No. of Pixels} & 123895 & 343332 & 123895 & 123895 & 123895 & 123895 & 123895 & 123895 & 123895 & 123895 \\
    \hline
    \hline
    \end{tabular}%
  \label{t1}%
\end{table*}%

\section{Comparison of Shadow Images by using numerical data analysis and the image processing}
\label{imageprocess}
This section presents image-based comparison of the Schwarzschild black hole shadow with the shadow of the JMN1 naked singularity to determine the difference/similarity between them. This comparison of the shadow images are performed on the basis of their intensity, contrast and structural properties. 

At the first instance, the Schwarzschild black hole shadow image is compared with the JMN1 naked singularity shadow images specifically on the basis of their intensities. Further, along with intensities, contrast and structural properties of the images are considered to precisely analyze the difference/similarity between the black hole and naked singularities shadows. MATLAB is utilized to analyze and create these shadow images from the theoretical intensity data which are acquired from the metrics of Schwarzschild black hole and JMN1 naked singularity. Initially, the comparison are performed for 20,000 intensity data points which are then increased to 2,00,000 points. To create the shadow images 20,000 and 2,00,000 intensity data points are interpolated to 2000 and 20,000 points, respectively. For the shadow image comparison, same number of pixels are set for all the images. The shadow images of the black hole and naked singularity for 2000 and 20,000 interpolated intensity data points are depicted in Fig.~(\ref{fig3}) and Fig.~(\ref{fig4}), respectively. It is observed that the quality of the shadow images are enhanced when the number of intensity data point are increased.

The difference/similarity of Schwarzschild black hole shadow with the shadow of the JMN1 naked singularity is determined initially by finding the arithmetic difference between their intensities and then some of the standard similarity metrics, namely normalized cross correlation (NCC), sum of absolute differences (SAD), and sum of squared differences (SSD), are employed along with the utilization of the Structural similarity (SSIM) index metric of MATLAB. The most general way to determine the degree of similarity between the two entities is defined by their correlation. The correlation coefficient is 1, if the two entities are exactly similar, whereas zero coefficient represents that the two entities are completely different. Both SAD and SSD metrics determines how close the entities are to each other. The SSD metric is much efficient in-terms of quality than the SAD metric, although the SAD is preferred over SSD metric because of its less computational requirement. However, in this section both the metric evaluation would be provided for the shadows comparison analysis. Further, to include other properties along with intensity, SSIM index metric is utilized. Being a good quality assessment tool, SSIM index is also the most suitable method to determine the difference/similarity between the images. The reason for selecting SSIM index metric for shadow analysis is that the metric assessment is based on three properties, viz., luminance, contrast, and structure. In digital images, luminance generally describes the intensity of emitted light, contrast is the difference between maximum and minimum pixel intensities in an image and structure provides the spatial information of the pixels. The SSIM index value, generally, range from 0 to 1. When the two images are exactly similar then the index value is 1 (White). In contrast, zero index (Black) value indicates that the two images under consideration are completely different. Thus, index value reduces as the similarity between images decreases. It should be noted that the SSIM index value could be negative for some image data types. The aforementioned similarity metrics would provide the insight into the degrees of difference by which the black hole shadow differs from the two naked singularities shadows. The following section presents the comprehensive comparative analysis of the black hole shadow image with the JMN1 shadow image for 20,000 (CASE-I) and 2,00,000 (CASE-II) intensity data points. For CASE-II, a more rigorous analysis is performed. Here, six different intensity data sets are generated to create the black hole and naked singularity shadow images from their respective metrics. Then the average images of the Schwarzschild black hole and the naked singularity shadow are created from each sets of shadow images. Thus, in CASE-II, the comparison analysis of the Schwarzschild black hole shadow with the naked singularity shadow are performed by utilizing their average images.

\subsection{Schwarzschild Black Hole and JMN1 Naked Singularity Shadow}
\label{ssec:1.A}

Initially, the arithmetic and absolute difference between the Schwarzschild black hole and JMN1 naked singularity shadow images for CASE-I (Figs.~(\ref{fig5in1}) and (\ref{fig5in2}) and CASE-II (Figs.~ (\ref{fig6in1}) and (\ref{fig6in2}) are obtained. Then the aforementioned similarity metrics are applied to achieve the in-depth comparative analysis. For CASE-I (CASE-II), the percentage difference with respect to the total number of considered pixels is achieved as 6.6532\% (1.8322\%), the correlation coefficient is 0.9962 (0.9998), the percentage difference relative to the Schwarzschild shadow image is 1.7672\% (0.2261\%), the percentage sum of squared difference is 0.7691\% (0.0418\%), and the SSIM index value is 0.9216 (0.9777). Fig.~(\ref{fig5in3}) and Fig.~(\ref{fig6in3}) represent the images achieved from the application of SSIM index metric for CASE-I and CASE-II, respectively. In these figures, the white color represents maximum similarity (SSIM index value is $\approx1$) region while the black color corresponds to minimum similarity (SSIM index value is much less than 1 even zero). 

The overall results of the comparison analysis between the two shadow images are presented in Table \ref{t1}. From the Table \ref{t1}, it is observed that with the increase of intensity data points the number of pixels increases which in-turn increases the resolution of the images. More intensity values would provide more information for the analysis but would also increase the size of images. Herein, for analyzing the shadows, images are cropped to maintain equal number of pixels. From the similarity metrics values and images it could be said that the increase in number of intensity data points makes the two shadow images appears to closely resemble with each other (see CASE-II). In other words, difference between Schwarzschild black hole and JMN1 naked singularity shadow for CASE-II is not apparent from the illustrated images. Hence, to represent the difference between them, a normalized histogram for the two cases are depicted in Fig.~(\ref{fig7}). This illustrations shows that even though the number of intensity data points are increased there still the exists a small difference between the two shadow images. Any further increase in the intensity data points would cause the percentage difference to reduce but still the two shadow images would not exactly match with each other.

\section{Conclusion}\label{result}
In this paper, we show that the JMN1 naked singularity and the Schwarzschild black hole can have different intensity distribution of light, though they have similar mass and shadow size. We define a parameter $\eta$ which shows the difference in the intensity of light in the image of a Schwarzschild black hole and JMN1 naked singularity, and it can be seen in Fig.~(\ref{fig3Eta}) that for $0 \leq b\leq 3\sqrt3$, the intensity of light in JMN1 spacetime is greater than that for Schwarzschild spacetime. However, two intensities $I_{0(JMN1)}$ and $I_{0 (SCH)}$ coincide with each other in the region $b> 3\sqrt3$. 

In addition, using the image processing technique, we show the difference in light intensity in Schwarzschild and JMN1 spacetimes. The following difference/similarity results are obtained when the Schwarzschild blackhole shadow image is compared with the JMN1 shadow image for 20,000 (2,00,000) data points: the percentage difference relative to the total number of considered pixels is obtained as $6.6532\%$ ($1.8322\%$), the correlation coefficient is 0.9962 (0.9998), the percentage difference relative to the Schwarzschild shadow image is $1.7672\%$ ($0.2261\%$), the percentage sum of squared difference is $0.7691\%$ ($0.0418\%$), and the SSIM index value is 0.9216 (0.9777). In addition, the normalized histogram illustrations are depicted for the two selected data points sets which shows that a small difference between the two shadow images would still exist even though the number of intensity data points is increased. The results which come out from the image processing technique are consistent with the theory. 

The differences in light intensity which are depicted here using image processing technique may be useful to analyse the observational data of the shadow image of a massive compact object. In future we plan to extend such an analysis to other blackhole and naked singularity spacetimes, and further enhancements in the results is expected with a higher computational power and adding more image analysis techniques. As of now, our results should be considered indicative, showing how a detailed analysis of the intrinsic properties of the shadows bring out the differences of the images, and how these can be used to decipher the physical and causal properties of the underlying object.


\end{document}